\documentclass[review]{elsarticle}

\usepackage[draft]{changes}
\usepackage{amsmath}
\usepackage{supertabular}
\usepackage{multirow}
\usepackage{amssymb}
\usepackage{amsmath}
\usepackage{graphicx}
\usepackage[normalem]{ulem}
\usepackage{longtable}
\usepackage{bm}
\usepackage{array}
\usepackage{etoolbox}
\usepackage{setspace}
\usepackage{tabularx}
\usepackage[draft]{changes}

\journal{Journal of \LaTeX\ Templates}

\begin{document}

\begin{frontmatter}

\title{Systematic study of $\mathcal{\alpha}$ decay half-lives based on Gamow--like model with a screened electrostatic barrier}


\author[mymainaddress]{Jun-Hao Cheng}
\author[mymainaddress]{Jiu-Long Chen}
\author[mymainaddress]{Jun-Gang Deng}
\author[mymainaddress2]{Xi-Jun Wu}%\corref{mycorrespondingauthor2}}
%\cortext[mycorrespondingauthor2]{Corresponding author}
\ead{wuxijun1980@yahoo.cn}
\author[mymainaddress,mymainaddress3,mymainaddress4]{Xiao-Hua Li\corref{mycorrespondingauthor1}}
\cortext[mycorrespondingauthor1]{Corresponding author}
\ead{lixiaohuaphysics@126.com}
\author[mymainaddress5]{Peng-Cheng Chu\corref{mycorrespondingauthor3}}
\cortext[mycorrespondingauthor3]{Corresponding author}
\ead{kyois@126.com }

\address[mymainaddress]{School of Nuclear Science and Technology, University of South China, 421001 Hengyang, People's Republic of China}
\address[mymainaddress2]{School of Math and Physics, University of South China, 421001 Hengyang, People's Republic of China}
\address[mymainaddress3]{Cooperative Innovation Center for Nuclear Fuel Cycle Technology $\&$ Equipment, University of South China, 421001 Hengyang, People's Republic of China}
\address[mymainaddress4]{Key Laboratory of Low Dimensional Quantum Structures and Quantum Control, Hunan Normal University, 410081 Changsha, People's Republic of China}
\address[mymainaddress5]{School of Science, Qingdao Technological University, 266000 Qingdao, People's Republic of China}

\begin{abstract}
In the present work we systematically study $\mathcal{\alpha}$ decay half-lives of $Z>51$ nuclei using the modified Gamow-like model which includes the effects of the centrifugal potential and electrostatic shielding. For the case of even-even nuclei, this model contains two adjustable parameters: the parameter $a$ related to the screened electrostatic barrier and the radius constant $r_0$, while for the case of odd-odd and odd-A nuclei, it is added a new parameter i.e. hindrance factor $h$ which is used to describe the effect of an odd-proton and/or an odd-neutron. Our calculations can well reproduce the experimental data. In addition, we use this modified Gamow-like model to predict the $\mathcal{\alpha}$-decay half-lives of seven even-even nuclei with $Z=120$ and some un-synthesized nuclei on their $\mathcal{\alpha}$ decay chains.
\end{abstract}


\end{frontmatter}

\section{Introduction}

$\mathcal{\alpha}$ decay, the spontaneous emission of a $^{4}$He by the nucleus and the formation of a new nuclides, was first defined by Rutherford in 1899. Since then, great efforts have been made in the realm of both theory and experiment, e.g., from the discovery of the atomic nucleus by $\mathcal{\alpha}$ scattering to the Geiger-Nuttall law describing a relationship between $\mathcal{\alpha}$ decay half-life and decay energy \cite{PhysRevC.85.044608, Dong2005, VIOLA1966741, PhysRevLett.103.072501}, from the barrier tunneling theory based on the quantum mechanics to the investigation of superheavy nuclei(SHN) \cite{Gamow1928,0034-4885-78-3-036301, RevModPhys.84.567, PhysRevLett.110.242502, SOBICZEWSKI2007292, RevModPhys.70.77, PhysRevLett.104.142502}. $\mathcal{\alpha}$ decay, as an important tool to investigate SHN, provides abundant information about the nuclear structure and stability of SHN\cite{0034-4885-78-3-036301, Hofmann2016, PhysRevC.85.044608, Yang2015}. Nowadays, there are many theoretical models used to study $\mathcal{\alpha}$ decay including the cluster model \cite{PhysRevLett.65.2975, PhysRevC.74.014304, XU2005303}, the unified model for $\mathcal{\alpha}$ decay and $\mathcal{\alpha}$ capture \cite{PhysRevC.73.031301, PhysRevC.92.014602}, the liquid drop model \cite{0305-4616-5-10-005, PhysRevC.48.2409, 0954-3899-26-8-305, PhysRevC.74.017304, GUO2015110}, the two-potential approach \cite{1674-1137-41-1-014102, PhysRevC.94.024338, PhysRevC.93.034316, PhysRevC.95.014319, PhysRevC.95.044303, PhysRevC.96.024318, PhysRevC.97.044322}, the empirical formulas \cite{PhysRevC.85.044608,0954-3899-39-1-015105, PhysRevC.80.024310, 0954-3899-42-5-055112} and others \cite{PhysRevLett.59.262, SANTHOSH201528, PhysRevC.87.024308,0954-3899-31-2-005, PhysRevC.81.064318, QI2014203}.

Recently, K. Pomorski \textit{et al.} proposed a Gamow-like model which is a simple phenomenological model based on the Gamow theory for the evaluations of half-life for $\mathcal{\alpha}$ decay \cite{PhysRevC.87.024308,1402-4896-2013-T154-014029}, while the nuclear potential is chosen as the square potential well, the centrifugal potential is ignored and the Coulomb potential is taken as the potential of a uniformly charged sphere with radius $R$ defined as Eq.(\ref{eq5}). They also extended this model to study the proton radioactivity\cite{Zdeb2016}, for the proton radioactivity shares the same mechanism as the $\mathcal{\alpha}$ decay. In 2016, Niu Wan \textit{et al.} systematically calculated the screened $\mathcal{\alpha}$ decay half-lives of the $\mathcal{\alpha}$ emitters with proton number $Z=52-105$ by considering the electrons in different external environments such as neutral atoms, a metal, and so on. They found that the decay energy and the interaction potential between $\mathcal{\alpha}$ particle and daughter nucleus are both changed due to the electrostatic shielding effect. And the electrostatic shielding effect is found to be closely related to the decay energy and its proton number\cite{Wan2016}. In 2017, R. Budaca and A. I. Budaca proposed a simple analytical model based on the WKB approximation for the barrier penetration probability which includes the centrifugal and overlapping effects besides the electrostatic repulsion\cite{Budaca2017}. In their model, there is only one parameter $a$ which is used to describe the electrostatic shielding effect of Coulomb potential by using the Hulthen potential\cite{Budaca2017}. They systematically calculated the half-lives of proton emission for $Z\geq 51$ nuclei. The results can well reproduce the experimental data.
Combining these points, in this work we modify the Gamow-like model proposed by K. Pomoski \textit{et al.}, considering the shielding effect of the Coulomb potential and the influence of the centrifugal potential, to systematically study the $\mathcal{\alpha}$ decay half-lives. All the database are taken from the latest atomic nucleus parameters from NUBASE 2016 \cite{1674-1137-41-3-030001}. We also extend our model to predict the $\mathcal{\alpha}$ decay half-lives of seven even-even superheavy nuclei with $Z=120$ and some un-synthesized nuclei on their $\mathcal{\alpha}$ decay chains.

This article is organized as follows. In Sec. II the theoretical framework for $\mathcal{\alpha}$ decay half-life is described in detail including Gamow-like model and other models such as Coulomb potential and Proximity potential model (CPPM) with Bass73 formalism, the Viola–Seaborg–Sobiczewski (VSS) empirical formula, the Universal curve (UNIV), Royer formula, the Universal decay law (UDL) and the Ni-Ren-Dong-Xu empirical formula (NRDX). In Sec. III, the detailed calculations, discussion and predictions are provided. A brief summary is given in Sec. IV.

\section{THEORETICAL FRAMEWORK}
\subsection{Gamow-like model}
$\mathcal{\alpha}$ decay half-life $T_\frac{1}{2}$, an important indicator of nuclear stability, can be calculated by the $\mathcal{\alpha}$ decay constant $\mathcal{\lambda}$ as
\begin{equation}
\label{1}
T_\frac{1}{2}=\frac{ln2}{\lambda}10^h,
\end{equation}
where $h$ is the so-called hindrance factor of $\mathcal{\alpha}$ decay due to the effect of an odd-proton and/or an odd-neutron. For the even-even nuclei, $h$ = 0, while for nuclei with an odd number of nucleons i.e. even-$N$, odd-$Z$ nuclei or odd-$N$, even-$Z$ $h=h_p=h_n$, odd-$N$, odd-$Z$ nuclei $2h=h_{np}$. The $\mathcal{\alpha}$ decay constant $\mathcal{\lambda}$ is given by \cite{PhysRevC.83.014601}
\begin{equation}
\label{2}
\lambda=\nu S_\alpha P,
\end{equation}
where $S_{\alpha}$ represents the preformation probability of $\mathcal{\alpha}$ particles in $\mathcal{\alpha}$ decay. According to Ref. \cite{PhysRevC.87.024308}, it can be known that the value of the preformation probability $S_{\alpha}$ can be changed by adjusting the radius constant $r_{0}$ appropriately. The results show that the best fitting result can be obtained with $S_{\alpha}$ =1, meanwhile $r_0\approx 1.2$fm, which also confirms the conclusion of Refs. \cite{PhysRevC.83.014601,0954-3899-17-S-045}. Then we choose $S_{\alpha}$=1 in this work.

$P$ given in Eq. (\ref{2}) represents the penetration probability of the $\mathcal{\alpha}$ particle crossing the barrier, calculated by the classical WKB approximation. Its concrete representation in the Gamow-like model is expressed as
\begin{equation}
\label{3}
P=\exp\! [- \frac{2}{\hbar} \int_{R}^{b} \sqrt{2\mu (V(r)-E_\text{k})}\, dr],
\end{equation}
here $E_\text{k}={Q_\alpha}{\frac{A-4}{A}}$ is the kinetic energy of $\mathcal{\alpha}$ particle emitted during $\mathcal{\alpha}$ decay. $Q_\alpha$ and $A$ are $\mathcal{\alpha}$ decay energy and the mass number of the parent nucleus, respectively. $b$ is the classical turning point. It satisfies the condition $V(b)=E_\text{k}$. $\mu=\frac{M_\text{d}M_{\alpha}}{(M_\text{d}+M_{\alpha})}$ is the reduced mass of the $\mathcal{\alpha}$ particle and the daughter nucleus in the center-of-mass coordinate with $M_{\text{d}}$ and $M_\alpha$ being masses of the daughter nucleus and $\mathcal{\alpha}$ particle. $V(r)$ is the total $\alpha$--daughter nucleus interaction potential.

In general, the $\alpha$--daughter nucleus electrostatic potential is by default of the Coulomb type as
\begin{equation}
\label{eq6}
V_C(r)=Z_{\alpha}Z_de^2/r,
\end{equation}
 where $Z_{\alpha}$ and $Z_d$ are the proton numbers of $\alpha$ particle and daughter nucleus. Whereas, in the process of $\alpha$ decay, for the superposition of the involved charges, movement of the emitted $\alpha$ particle which generates a magnetic field and the inhomogeneous charge distribution of the nucleus, the emitted $\alpha$-daughter nucleus electrostatic potential behaves as a Coulomb potential at short distance and drop exponentially at large
distance i.e. the screened electrostatic effect\cite{Budaca2017}. This behavior of electrostatic potential can be described as the Hulthen type potential which is widely used in nuclear, atomic, molecular and solid state physics\cite{doi:10.1063/1.4995175, PhysRevC.91.034614} and defined as
\begin{equation}
\label{eq7}
V_{h}(r)=\frac{a Z_\text{d} Z_\alpha e^2}{e^{ar}-1},
\end{equation}
where $a$ is the screening parameter. In this framework, the total $\alpha$--daughter nucleus interaction potential $V(r)$ is given by
\begin{equation}
V(r)=\left\{
\begin{array}{rcl}
-V_0,& & {0 \leq r \leq R,}\\
V_h(r)+V_l(r), & & {r \ge R,}
\end{array} \right.
\end{equation}
where $V_0$ is the depth of the square well. $V_h(r)$ and $V_l(r)$ are the Hulthen type of screened electrostatic Coulomb potential and centrifugal potential, respectively. The spherical square well radius $R$ is equal to the sum of the radii of both daughter nucleus and $\mathcal{\alpha}$ particle, it is expressed as
\begin{equation}
\label{eq5}
R=r_0({A_\text{d}}^\frac{1}{3}+{A_\alpha}^\frac{1}{3}),
\end{equation}
where $A_{\text{d}}$ and $A_{\alpha}$ are the mass number of the daughter nucleus and $\mathcal{\alpha}$ particle, respectively. $r_{0}$, the radius constant, is the adjustable parameter in our model.

Because $l(l + 1) \rightarrow (l + 1/2)^2$ is a necessary correction for one-dimensional problem \cite{Gur31}, the centrifugal potential $V_{l}(r)$ is chose as the Langer modified form in this work. It can be expressed as
\begin{equation}
\label{eq10}
V_{l}(r)=\frac{\hbar^2(l+\frac{1}{2})^2}{2{\mu}r^2},
\end{equation}
where $l$ is the orbital angular momentum taken away by the $\mathcal{\alpha}$ particle. $l = 0$ for the favored $\mathcal{\alpha}$ decays, while $l\ne 0$ for the unfavored decays. Based on the conservation laws of party and angular momentum \cite{PhysRevC.82.059901}, the minimum angular momentum $l_{\text{min}}$ taken away by the $\mathcal{\alpha}$ particle can be determined by
\begin{equation}
\
l_{\text{min}}=\left\{\begin{array}{llll}

{\Delta}_j,&\text{for even${\Delta}_j$ and ${\pi}_p$= ${\pi}_d$},\\

{\Delta}_j+1,&\text{for even${\Delta}_j$ and ${\pi}_p$$\ne$${\pi}_d$},\\

{\Delta}_j,&\text{for odd${\Delta}_j$ and ${\pi}_p$$\ne$${\pi}_d$},\\

{\Delta}_j+1,&\text{for odd${\Delta}_j$ and ${\pi}_p$= ${\pi}_d$},

\end{array}\right.
\label{6}
\end {equation}
where ${\Delta}_j= |j_p-j_d|$. $j_p$, ${\pi}_p$, $j_d$, ${\pi}_d$ represent spin and parity values of the parent and daughter nuclei, respectively.

The $\nu$ represents the collision frequency of $\mathcal{\alpha}$ particle in the potential barrier. It can be calculated with the oscillation frequency $\omega$ and expressed as \cite{PhysRevC.81.064309}
\begin{equation}
\label{9}
\nu=\omega/2\pi=\frac{(2n_\text{r}+l+\frac{3}{2})\hbar}{2\pi \mu {R_\text{n}}^2}=\frac{(G+\frac{3}{2})\hbar}{1.2\pi \mu{R_\text{0}}^2},
\end{equation}
where $R_\text{n}= \sqrt{3/5}R_\text{0}$ is the nucleus root-mean-square (rms) radius and $R_0 = 1.28A^{1/3}-0.76+0.8A^{-1/3}$ is the radius of the parent nucleus. $G=2n_\text{r}+l$ is the main quantum number with $n_\text{r}$ and $l$ being the radial quantum number and the angular quantity quantum number, respectively. In the work of Ref. \cite{PhysRevC.69.024614}, for $\mathcal{\alpha}$ decay, $G$ can be obtained by
\begin{equation}
G=2n_\text{r}+l=\left\{
\begin{array}{rcl}
18, & & {N \leq 82,}\\
20, & & {82 < N \leq 126,}\\
22, & & {N > 126.}
\end{array} \right.
\end{equation}

\subsection{Other models}
\subsubsection{Coulomb potential and Proximity potential model with proximity potential Bass73 formalism (CPPM-Bass73)}
In CPPM, the $\mathcal{\alpha}$ decay half-life $T_\frac{1}{2}$ is related to the decay constant $\mathcal{\lambda}$ as
\begin{equation}
T_\frac{1}{2}=\frac{ln2}{\lambda},
\end{equation}
where the decay constant $\mathcal{\lambda}$ can be obtained by
\begin{equation}
\lambda=\nu P.
\end{equation}
The assault frequency $\nu$ can be calculated with the oscillation frequency $\omega$ and expressed as 
\begin{equation}
\nu=\omega/2\pi=2E_v/h,
\end{equation}
where $h$ is the Planck constant. The zero-point vibration energy $E_v$ can be calculated with $Q_\alpha$ and expressed as \cite{Poenaru1986}
\begin{equation}
\
E_v=\left\{\begin{array}{llll}

0.1045Q_\alpha,\text{for even-even nuclei},\\

0.0962Q_\alpha,\text{for even-$N$, odd-$Z$ nuclei},\\

0.0907Q_\alpha,\text{for odd-$N$, even-$Z$ nuclei},\\

0.0767Q_\alpha,\text{for odd-odd nuclei}.

\end{array}\right.
\end {equation}
$P$ denote the semiclassical WKB barrier penetration probability, which is expressed as
\begin{equation}
P=\exp\! [- \frac{2}{\hbar} \int_{R_{in}}^{R_{out}} \sqrt{2\mu (V(r)-E_\text{k})}\, dr],
\end{equation}
where $R_{in}$ and $R_{out}$ are the classical turning points which satisfy the conditions $V(R_{in})=V(R_{out})=Q_\alpha$. The total interaction potential $V (r)$, between the emitted proton and daughter nucleus, including nuclear, Coulomb and centrifugal potential barriers. It can be expressed as
\begin{equation}
V(r)=V_N(r)+V_C(r)+V_l(r)
\end{equation}
$V_l(r)$ are same as Eq.(\ref{eq10}), $V_C(r)$ can be expressed as

\begin{equation}
\
V_C(r)=\left\{\begin{array}{ll}

\frac{Z_{\alpha}Z_de^2}{2R}[3-(\frac{r}{R})^2],&r<R,\\

\frac{Z_{\alpha}Z_de^2}{r},&r>R.

\end{array}\right.
\label{subeq:1}
\end {equation}

We select proximity potential Bass73 to calculate the nuclear potential $V_N(r)$ \cite{BASS1973139, BASS197445}, which is given by
\begin{equation}
V_N(r)=-4\pi\gamma\frac{dR_1R_2}{R}exp(-\frac{\xi}{d})=\frac{-da_s{A_\text{d}^\frac{1}{3}}{{A_\alpha}^\frac{1}{3}}}{R}exp(-\frac{r-R}{d}),
\end{equation}
where the $d=1.35$ fm is the range parameter, and the surface term in the liquid drop model mass formula $a_s=17.0$ MeV. The $\gamma$ is the specific surface energy of the liquid drop model. $R=r_0({A_\text{d}}^\frac{1}{3}+{A_\alpha}^\frac{1}{3})$ represents the the sum of the half-maximum density radii with $r_0 = 1.07$ fm.

\subsubsection{The Viola–Seaborg–Sobiczewski (VSS) semi-empirical relationship}
The Viola–Seaborg–Sobiczewski semi-empirical relationship, one of the commonly used formulas for calculating the half-life of $\alpha$ decay, is proposed by Viola and Seaborg and the value given by Sobiczewski instead of the original value given by Viola and Seaborg \cite{VIOLA1966741}. It can be expressed as
\begin{equation}
log_{10}(T_\frac{1}{2})=(aZ+b)Q^{-1/2}+cZ+d+h_{log},
\end{equation}
where $Z$ is the atomic number of the parent nucleus and $h_{log}$ is hindrance factor. The values of parameters are $a = 1.66175, b=-8.5166, c=-0.20228, d=-33.9069$ and
\begin{equation}
\
h_{log}=\left\{\begin{array}{llll}

0       ,&\text{for even-even nuclei},\\

0.772,&\text{for even-$N$, odd-$Z$ nuclei},\\

1.066,&\text{for odd-$N$, even-$Z$ nuclei},\\

1.114,&\text{for odd-odd nuclei}.

\end{array}\right.
\end {equation}

\subsubsection{The Universal curve (NUIV)}
Poenaru et al. proposed the Universal (UNIV) curve for calculating the decay half-lives by extending a fission theory to larger asymmetry, which can be expressed as \cite{PhysRevC.83.014601, PhysRevC.85.034615}
\begin{equation}
log_{10}T_\frac{1}{2}=-log_{10}P-log_{10}S_\alpha+[log_{10}(ln2)-log_{10}\nu].
\end{equation}
The penetrability of an external Coulomb barrier $P$ may be obtained analytically as\cite{SANTHOSH201833}
\begin{equation}
-log_{10}P=0.22873\sqrt{\mu Z_d{Z_\alpha}R_b} \times [arccos {\sqrt{r}}-\sqrt{(r(1-r))}],
\end{equation}
where $r=R_a/R_b$ fm with $R_a=1.2249({A_\text{d}}^\frac{1}{3}+{A_\alpha}^\frac{1}{3})$ fm and $R_b=1.43998Z_d{Z_\alpha}/Q_\alpha$ fm being the two classic turning points. The logarithmic form of  the pre-formation factor is given by
\begin{equation}
log_{10}S_\alpha=-0.598(A_\alpha-1).
\end{equation}
$C=[-log_{10}\nu+log_{10}(ln2)]=-22.16917$ is the additive constant \cite{PhysRevC.83.014601, PhysRevC.85.034615}.

\subsubsection{Royer formula}

Royer proposed the analytical formula for determining $\alpha$ decay half-lives by fitting $\alpha$ emitters experimental data \cite{0954-3899-26-8-305}. It can be written as
\begin{equation}
log_{10}T_\frac{1}{2}=a+bA^{1/6}Z^{1/2}+\frac{cZ}{{Q_\alpha}^{1/2}}.
\end{equation}
The parameters a, b and c are given by
\begin{equation}
\
\left\{\begin{array}{llll}

a=-25.31,b=-1.1629,c=1.5864,\text{for even-even nuclei},\\

a=-25.68,b=-1.1423,c=1.5920,\text{for even-$N$, odd-$Z$ nuclei},\\

a=-26.65,b=-1.0859,c=1.5848,\text{for odd-$N$, even-$Z$ nuclei},\\

a=-29.48,b=-1.1130,c=1.6971,\text{for odd-odd nuclei}.

\end{array}\right.
\end {equation}

\subsubsection{The Universal decay law (UDL)}
Qi et al. given a new universal decay law (UDL) for describing $\alpha$-decay and cluster decay modes starting from $\alpha$-like $R$-matrix theory and the microscopic mechanism of the charged-particle emission\cite{PhysRevLett.103.072501, PhysRevC.80.044326}. It can be expressed as

\begin{equation}
log_{10}T_\frac{1}{2}=a\chi'+b\rho'+c,
\end{equation}
where $\chi'=a{Z_\alpha}Z_d\sqrt{\frac{\mu}{Q_\alpha}}$ and $\rho'=\sqrt{A{Z_\alpha}Z_d({A_\text{d}}^\frac{1}{3}+{A_\alpha}^\frac{1}{3})}$. Here the parameters $a=0.4314, b=-0.4087$ and $c=-25.7725$ are determined by fitting to experiments of $\alpha$ and cluster decays.\cite{PhysRevLett.103.072501, PhysRevC.80.044326}

\subsubsection{The Ni-Ren-Dong-Xu empirical formula (NRDX)}
Ni et al. proposed a new general formula with three parameters for determining half-lives and decay energies of $\alpha$ decay and cluster radioactivity \cite{PhysRevC.78.044310}.This new formula is directly deduced from the WKB barrier penetration probability with some approximations. Their calculations by using this formula show excellent agreement between the experimental data and the calculated values. It can be given by,
\begin{equation}
log_{10}T_\frac{1}{2}=a\sqrt{\mu}{Z_\alpha}Z_d{Q}^{-1/2}+b\sqrt{\mu}(Z_dZ_\alpha)^{1/2}+c,
\end{equation}
The parameters a, b and c are given by
\begin{equation}
\
\left\{\begin{array}{lll}

a=0.39961,\\

b=-1.31008,\\	

c_{e-e}=-17.00698.

\end{array}\right.
\end {equation}
This formula successfully combines the phenomenological laws of $\alpha$ decay and cluster radioactivity.

\section{RESULTS AND DISCUSSION}

In this work, we use the least squares principle to fit the adjustable parameters, while the database are taken from the latest evaluated nuclear properties table NUBASE2016 \cite{1674-1137-41-3-030001}. At first, for the parameter $h$ being used to describe the effect of an odd-proton and/or an odd-neutron, we choose the experimental data of $\mathcal{\alpha}$ decay half-lives of 169 even-even nuclei as the database to obtain the parameters $a$ and $r_0$, while $h=0$. Then choosing the experimental data of $\mathcal{\alpha}$ decay half-lives of 132 odd-$N$, even-$Z$ nuclei, 94 even-$N$, odd-$Z$ nuclei and 66 doubly-odd nuclei as the database to determine the parameter $h$, while fixed the parameters $a$ and $r_0$, using the relationship $h_n=h_p=\frac{1}{2}h_{np}=h$. The values of 3 adjustable parameters are given as
\begin{equation}
\ r_0=1.14\text{fm}, a=7.8\times10^{-4},\\
\ h=0.3455.
\end{equation}

\begin{figure}
\includegraphics[width=9.0cm]{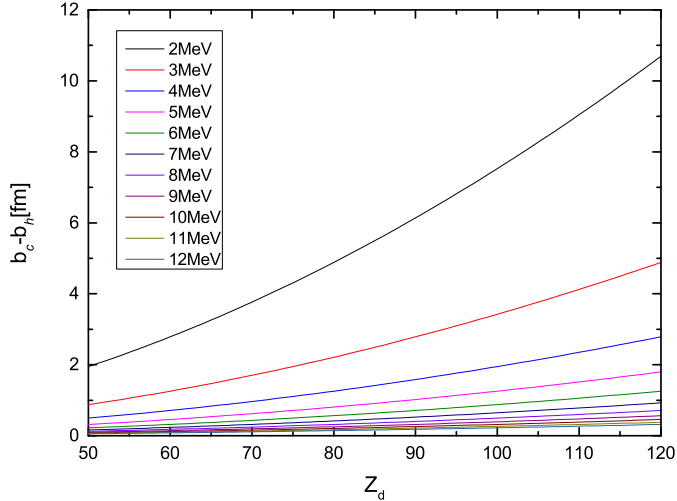}
\caption{The difference between $b_c$ and $b_h$ obtained by $V(r)=E_\text{k}$ only considering the Coulomb potential. For obtained $b_c$, Coulomb potential is taken as the potential of a uniformly charged sphere expressed as Eq. (\ref{eq6}), while for $b_h$ Coulomb potential is taken as Hulthen potential with $a=7.8\times10^{-4}$ expressed as Eq. (\ref{eq7}).}
\label{fig6}
\end{figure}

The value of $a$ is small but it observably impacts on the classical turning point $b$, whereas the $\mathcal{\alpha}$ decay half-life is sensitive to $b$. For intuitively display the effects, in Fig. \ref{fig6} we show the different kinetic energy $E_\text{k}$ values correspond to difference in $b$ values for the pure Coulomb and Hulthen potential, i.e., no centrifugal potential contribution, where $b_c$ and $b_h$ represent the $b$ value calculated using Coulomb and using the Hulthen potential, respectively. From this figure, we can find that the smaller decay energy and larger proton number of the daughter nucleus are, the greater difference in the $b$ value between the pure Coulomb and the Hulthen potential be.

\begin{figure}
\includegraphics[width=14.0cm]{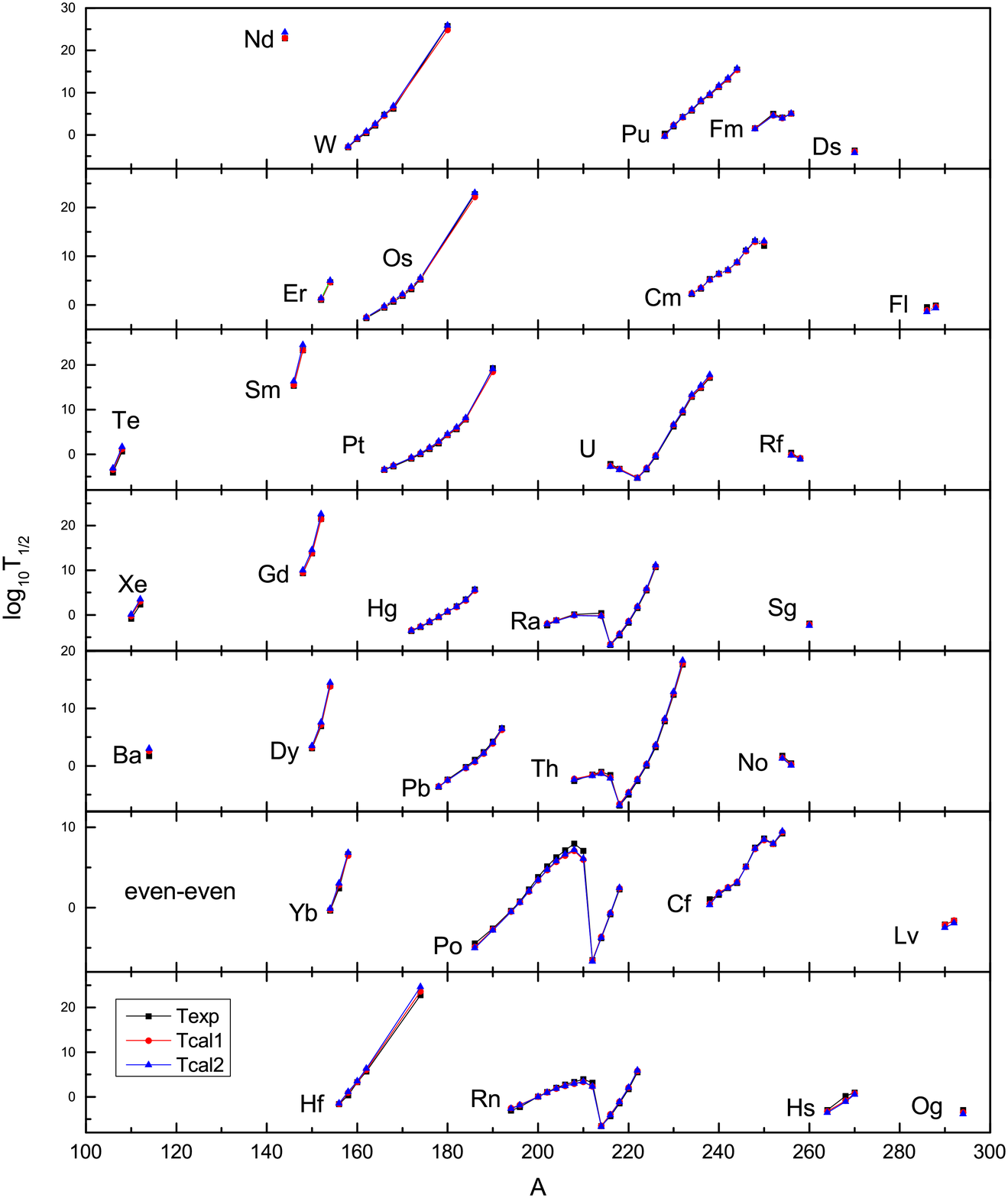}
\caption{The calculation of the $\mathcal{\alpha}$ decay half-life of the even-even nuclei. ${\text{lg}T_{1/2}^{\text{cal1}}}$ is the logarithmic form of the $\mathcal{\alpha}$ decay half-life calculated in this work, and ${\text{lg}T_{1/2}^{\text{cal2}}}$ is the logarithmic form of the $\mathcal{\alpha}$ decay half-life calculated by the theoretical model and parameters in Ref.\cite{PhysRevC.87.024308}. The experimental $\mathcal{\alpha}$ decay half-lives and decay energies are taken from the latest evaluated nuclear properties table NUBASE2016 \cite{1674-1137-41-3-030001} and evaluated mass number table AME2016 \cite{1674-1137-41-3-030003}. }

\label{fig7}
\end{figure}

\begin{figure}
\includegraphics[width=14.0cm]{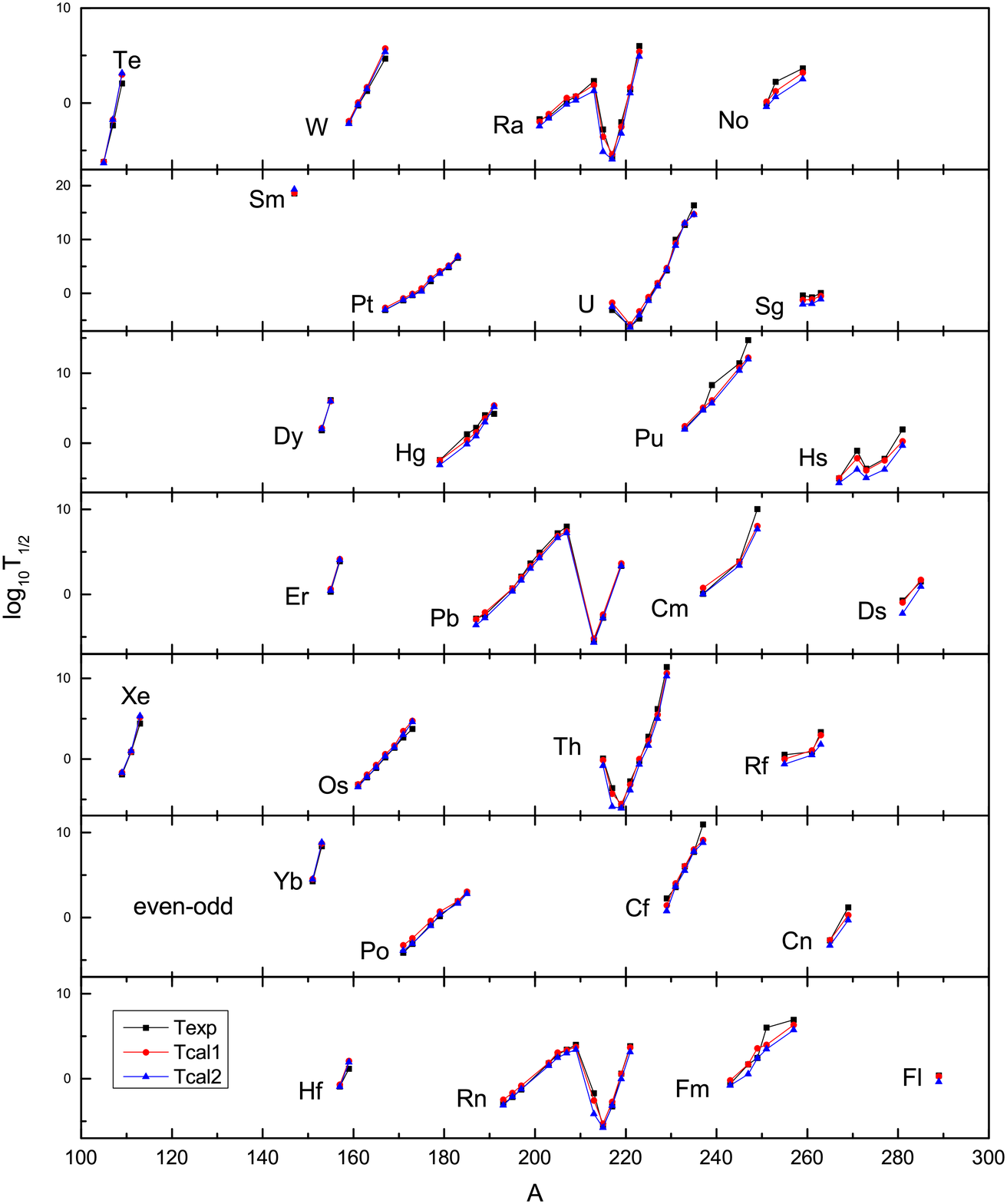}
\caption{The same as Fig. \ref{fig7}, but for the case of even-$Z$, odd-$N$ nuclei.}

\label{fig8}
\end{figure}

\begin{figure}
\includegraphics[width=14.0cm]{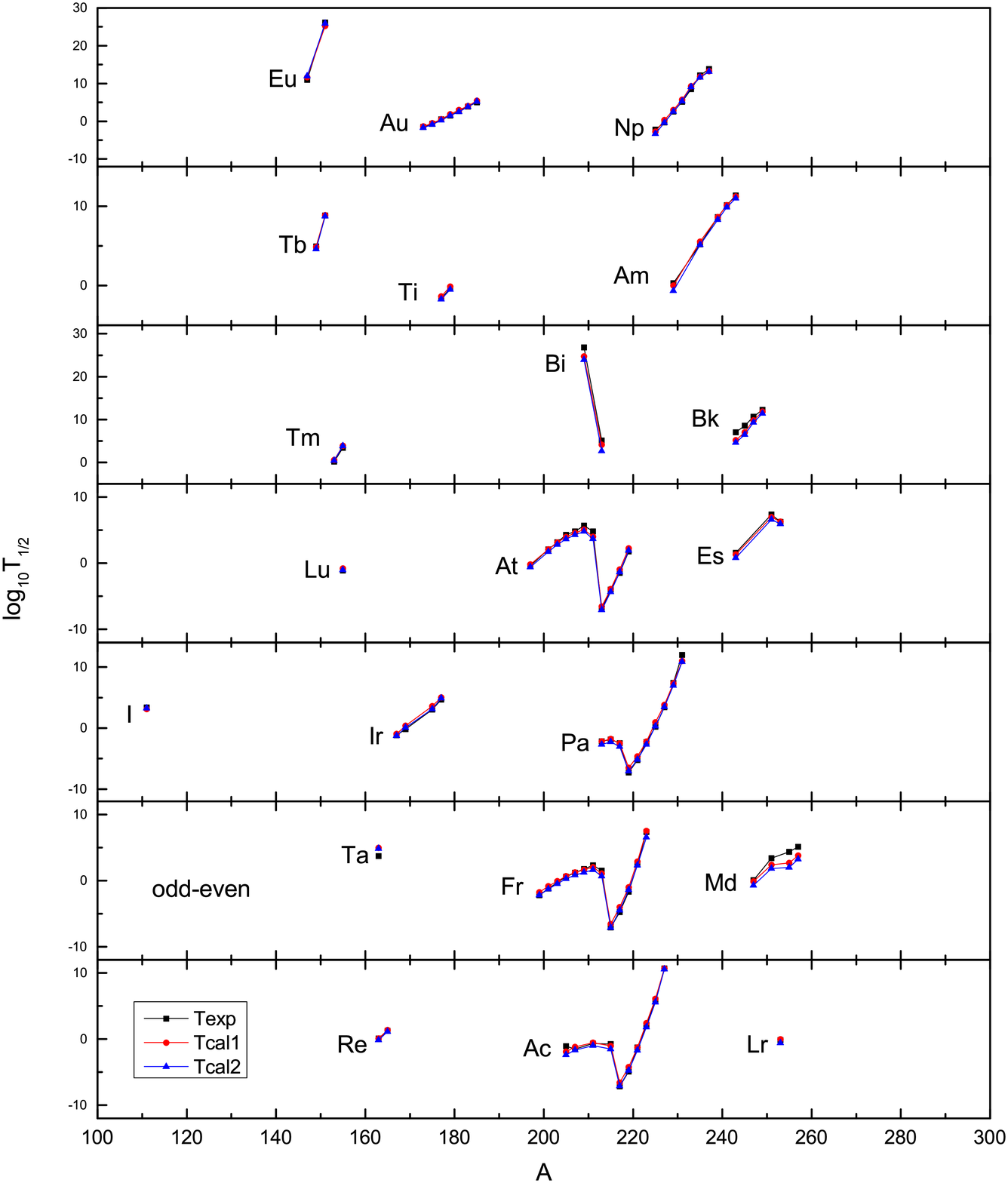}
\caption{The same as Fig. \ref{fig7}, but for the case of odd-$Z$, even-$N$ nuclei.}

\label{fig9}
\end{figure}

\begin{figure}
\includegraphics[width=14.0cm]{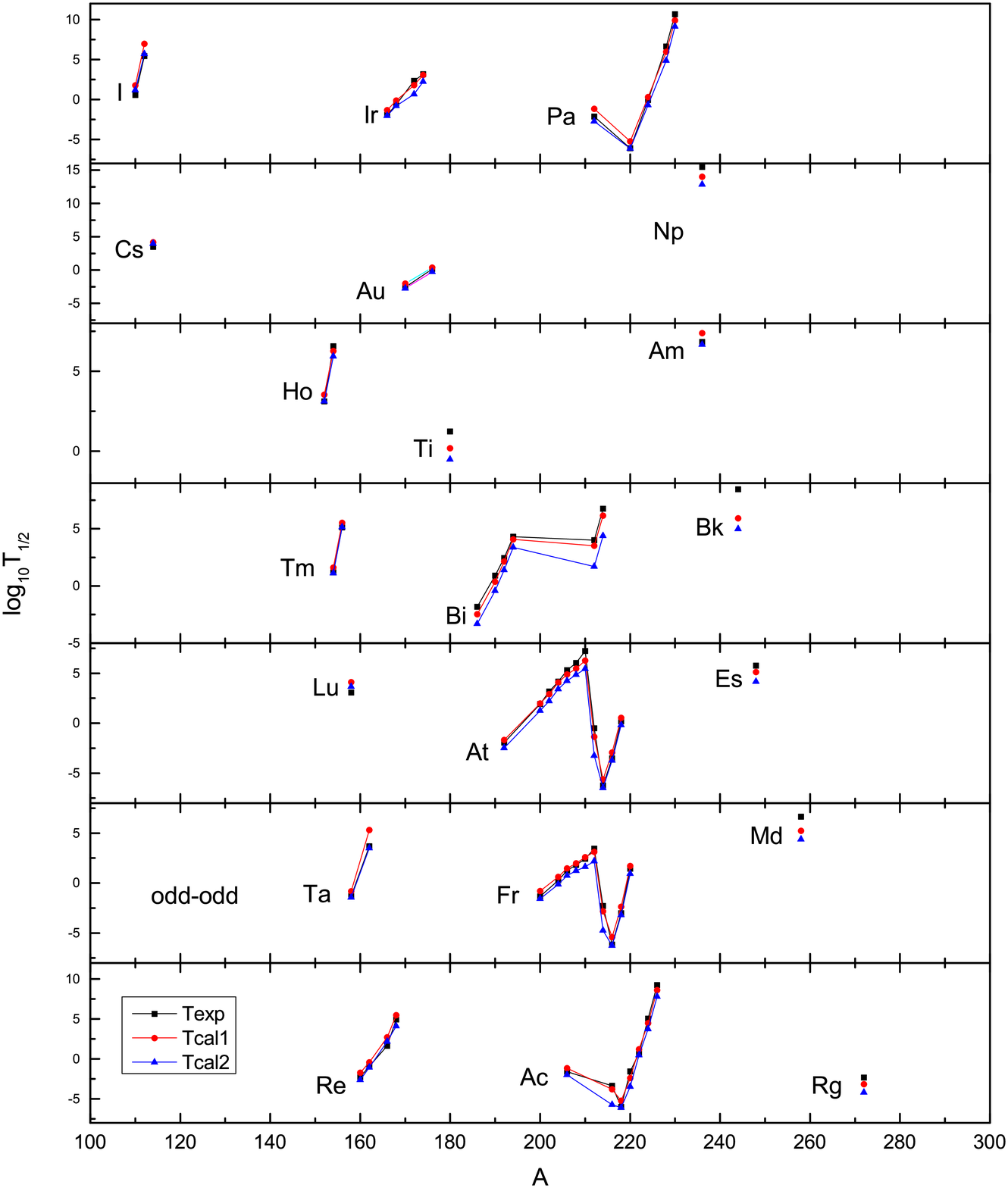}
\caption{The same as Fig. \ref{fig7}, but for the case of odd-$Z$, odd-$N$ nuclei.}

\label{fig10}
\end{figure}

Using our modified Gamow--like model, we systematically calculate the $\mathcal{\alpha}$ decay half-lives of even-even, odd-odd, odd-$A$ nuclei. The detailed results are shown in the Fig. \ref{fig7} -- \ref{fig10}. In Fig. \ref{fig7}, we show the 169 $\mathcal{\alpha}$ decay experimental data of even-even nuclei and the theoretical values of $\mathcal{\alpha}$ decay half-live calculated by different methods. The X-axis represents the mass number in the corresponding $\mathcal{\alpha}$ decay, the Y-axis represents the logarithmic of the $\mathcal{\alpha}$ decay half-life. The three coordinate points represent logarithmic form of the experimental $\mathcal{\alpha}$ decay half-lives, logarithmic forms of the calculated $\mathcal{\alpha}$ decay half-lives in this work denoted as ${\text{lg}T_{1/2}^{\text{cal1}}}$ and by the theoretical model and parameters in Ref. \cite{PhysRevC.87.024308} denoted as ${\text{lg}T_{1/2}^{\text{cal2}}}$, respectively. The cases of even-$Z$, odd-$N$ nuclei, odd-$Z$, even-$N$ nuclei and odd-$N$, odd-$Z$ nuclei are shown in Fig. \ref{fig8}, Fig. \ref{fig9} and Fig. \ref{fig10}, respectively. The meanings of each coordinate in Fig. \ref{fig8} -- \ref{fig10} is same as Fig. \ref{fig7}.

As can be seen from the Fig. \ref{fig7} -- \ref{fig10}, the ${\text{lg}{T^{\text{cal1}}_{1/2}}}$ can better reproduce with experimental data than ${\text{lg}{T^{\text{cal2}}_{1/2}}}$. In order to intuitively compare ${{T^{\text{cal1}}_{1/2}}}$ with ${{T^{\text{cal2}}_{1/2}}}$, we calculate the standard deviation $\sigma=\sqrt{\sum ({\text{lg}{T^{\text{expt}}_{1/2}}}-{\text{lg}{T^{\text{cal}}_{1/2}}})^2/n}$ between $\mathcal{\alpha}$ decay half-lives of calculations and experimental data. The results $\sigma_1$, $\sigma_2$ represent standard deviations between ${\text{lg}{T^{\text{cal1}}_{1/2}}}$, ${\text{lg}{T^{\text{cal2}}_{1/2}}}$ and ${\text{lg}{T^{\text{expt}}_{1/2}}}$, which are given in the Table \ref{table 6}. From this table, we can clearly see that for the cases of even-even, odd-proton, odd-neutron and doubly-odd nuclei, our calculations ${\text{lg}{T^{\text{cal1}}_{1/2}}}$ improve $\frac{0.487-0.348}{0.487}\approx28.5\%$, $\frac{0.967-0.681}{0.967}\approx29.6\%$, $\frac{0.789-0.598}{0.789}\approx24.2\%$ and $\frac{1.235-0.748}{1.235}\approx39.4\%$ compared to ${\text{lg}{T^{\text{cal2}}_{1/2}}}$, respectively. It is shown that ${T^{\text{cal1}}_{1/2}}$ can better reproduce with experimental data than ${T^{\text{cal2}}_{1/2}}$ by considering the shielding effect of the Coulomb potential and the centrifugal potential in this work. And for even-even nuclei, ${T^{\text{cal2}}_{1/2}}$ are calculated by Gamow-like model proposed by K. Pomorski \cite{PhysRevC.87.024308} which contain only one parameter $r_0$, and ${T^{\text{cal1}}_{1/2}}$ are calculated by our improved Gamow-like model with two parameters $r_0$ and $a$. So the addition of the parameter $a$ makes the ${T^{\text{cal1}}_{1/2}}$ more consistent with the experimental data than ${T^{\text{cal2}}_{1/2}}$. In many of the Fig. \ref{fig7} -- \ref{fig10} we can see dips and peaks in half-lives, it is because $Z/N=50$, $Z/N=82$ and $N=126$ is the magic core, and the nucleons in the core play an essential role on the $\alpha$ preformation probability\cite{PhysRevC.94.024338}.

\begin{table}
\caption{Compare root-mean-square deviations of $\text{lg}T_{1/2}$ between our calculations and calculations using parameters and models of Ref. \cite{PhysRevC.87.024308}. In the first row of the table, ${\pi_\text{z}}$ and ${\pi_\text{n}}$ are parity of the number of protons and neutrons, respectively. The second row is the corresponding total number of nuclei, and the third row is the corresponding $h$ value. The fourth row is the root-mean-square of this work, and the fifth row is the root-mean-square of Ref.\cite{PhysRevC.87.024308}}
\label{table 6}

\begin{tabular}{ccccc}

\multicolumn{1}{c}{${\pi_\text{z}}-{\pi_\text{n}}$}&n&$h$&$\sigma_1$&$\sigma_2$\\
\hline
\\
e-e&169&{--}&{0.348}&{0.487}\\
\\
e-o&132&0.3455&{0.681}&{0.967}\\
\\
o-e&94&0.3455&{0.598}&{0.789}\\
\\
o-o&66&0.691&{0.748}&{1.235}\\
\hline
\end{tabular}
\end{table}

The synthesis and research of SHN have became a hot topic in nuclear physics \cite{PhysRevC.97.064609, 1674-1137-41-7-074106, PhysRevC.98.014618}. Now we extend our model to predict the $\alpha$ decay half-lives of nuclei $Z=120$ i.e. $^{296}120$, $^{298}120$,$^{300}120$,$^{302}120$,$^{304}120$,$^{306}120$ as well as $^{308}120$ and some un-synthesized nuclei on their $\mathcal{\alpha}$ decay chains. From the conclusion of decay properties for SHN in Ref. \cite{SANTHOSH201833}, we can obtain the $\mathcal{\alpha}$ decay chains of these nuclei, which are $^{296}120\to^{292}\text{Og}\to^{288}\text{Lv}\to^{284}\text{Fl}\to^{280}\text{Cn}\to^{276}\text{Ds}\to^{272}\text{Hs}\to^{268}\text{Sg}$, $^{298}120\to^{294}\text{Og}\to^{290}\text{Lv}\to^{286}\text{Fl}\to^{282}\text{Cn}\to^{278}\text{Ds}\to^{274}\text{Hs}$, $^{300}120\to^{296}\text{Og}\to^{292}\text{Lv}\to^{288}\text{Fl}\to^{284}\text{Cn}$, $^{302}120\to^{298}{\text{Og}}\to^{294}\text{Lv}\to^{290}\text{Fl}$, $^{304}120\to^{300}\text{Og}\to^{296}\text{Lv}\to^{294}\text{Fl}$, $^{306}120\to^{302}\text{Og}\to^{298}\text{Lv}$, and $^{308}120\to^{304}\text{Og}\to^{300}\text{Lv}$. In our previous studies of the superheavy nucleus\cite{1674-1137-42-4-044102,1674-1137-41-12-124109}, the $\mathcal{\alpha}$ decay energy is one key input for calculating the $\mathcal{\alpha}$ decay half-life. Meanwhile Sobiczewski \cite{PhysRevC.94.051302} discovered that the calculation taking $\mathcal{\alpha}$ decay energy from WS3+ \cite{PhysRevC.84.051303} can best reproduce experimental $\mathcal{\alpha}$ decay half-life. In the present work, we use $\mathcal{\alpha}$ decay energy from WS3+ to calculate the half-life of even-even nuclide with proton number $Z=120$ and nuclei on their $\mathcal{\alpha}$ decay chains except the five known nuclei i.e. $^{294}$Og, $^{290}$Lv, $^{286}$Fl, $^{292}$Lv and $^{288}$Fl are taken from NUBASE2016 \cite{1674-1137-41-3-030001}.

\begin{figure}
\begin{minipage}{0.48\linewidth}
 \centerline{\includegraphics[height=5.0cm,width=6.0cm]{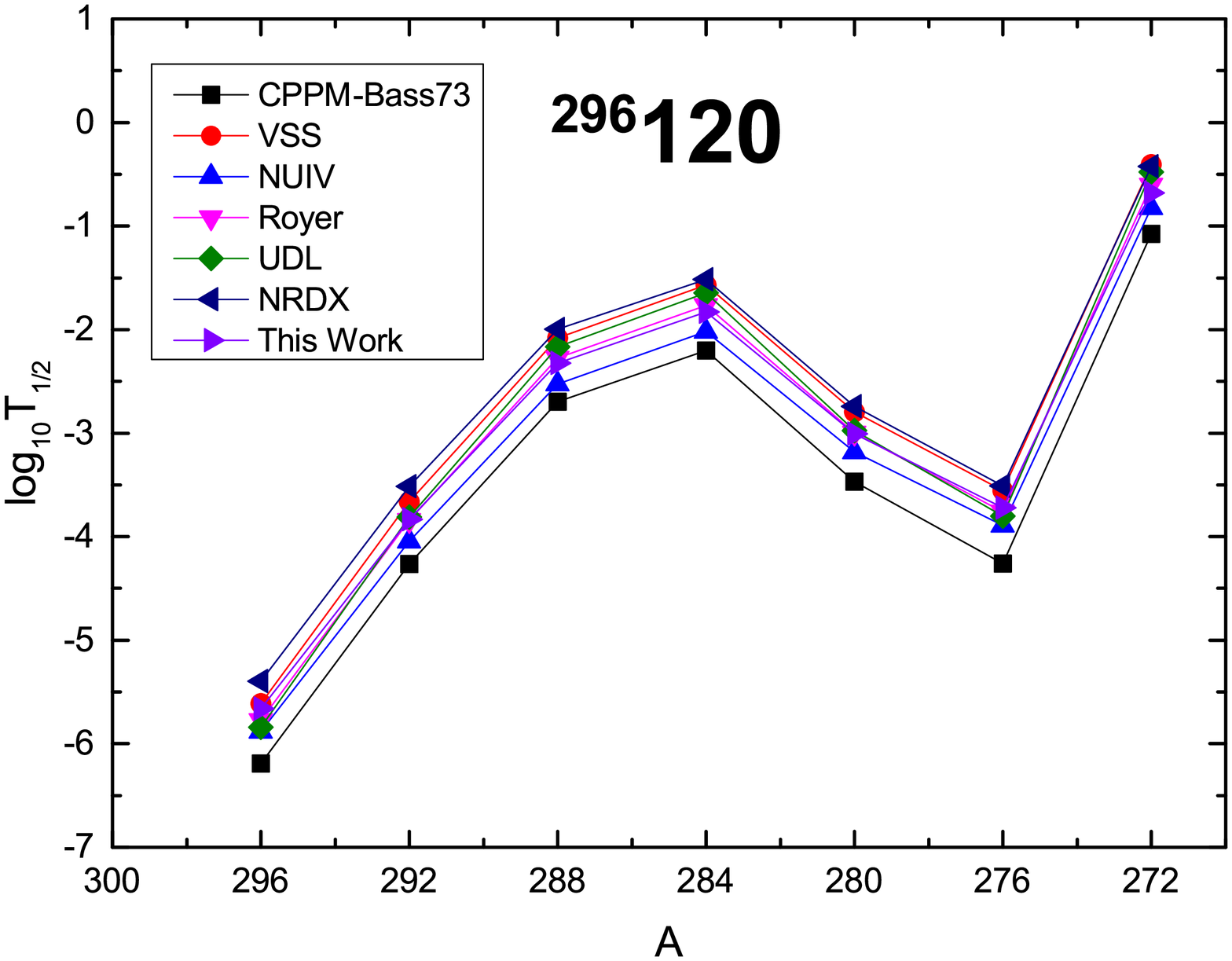}}
 \centerline{The case of $ ^{296}120$}
\end{minipage}
\hfill
\begin{minipage}{0.48\linewidth}
 \centerline{\includegraphics[height=5.0cm,width=6.0cm]{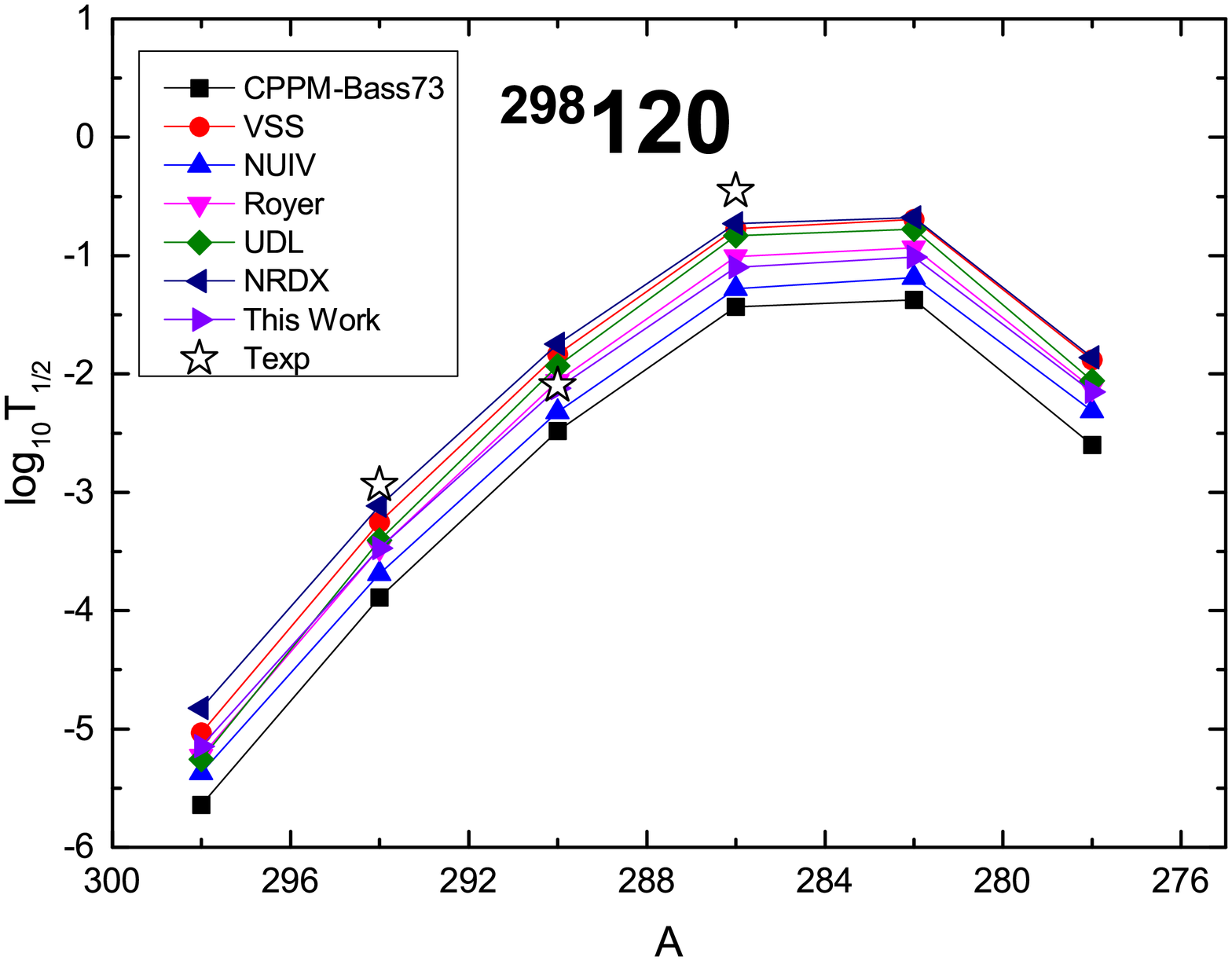}}
 \centerline{The case of $ ^{298}120$}
\end{minipage}
\vfill
\begin{minipage}{.48\linewidth}
 \centerline{\includegraphics[height=5.0cm,width=6.0cm]{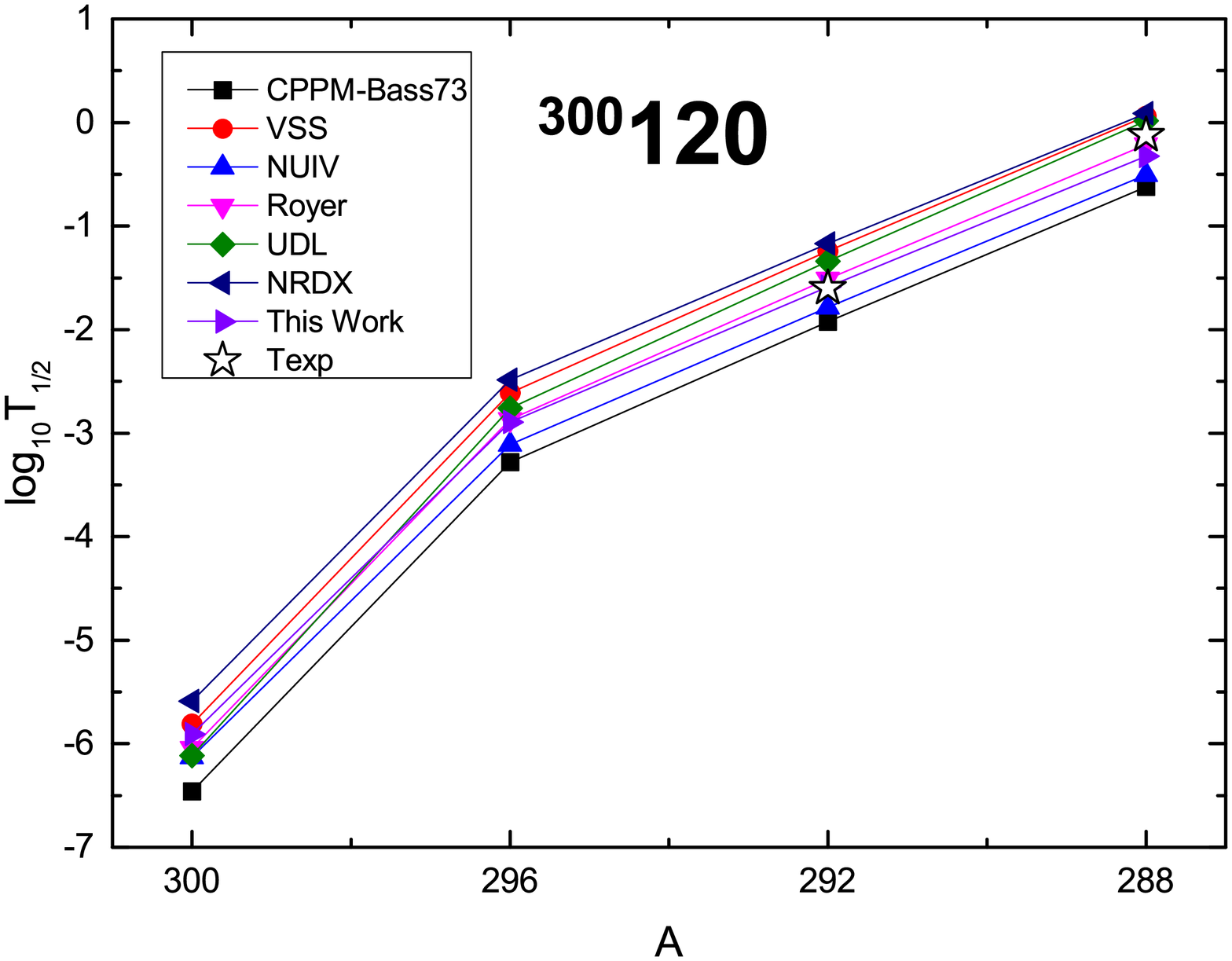}}
 \centerline{The case of $ ^{300}120$}
\end{minipage}
\hfill
\begin{minipage}{0.48\linewidth}
 \centerline{\includegraphics[height=5.0cm,width=6.0cm]{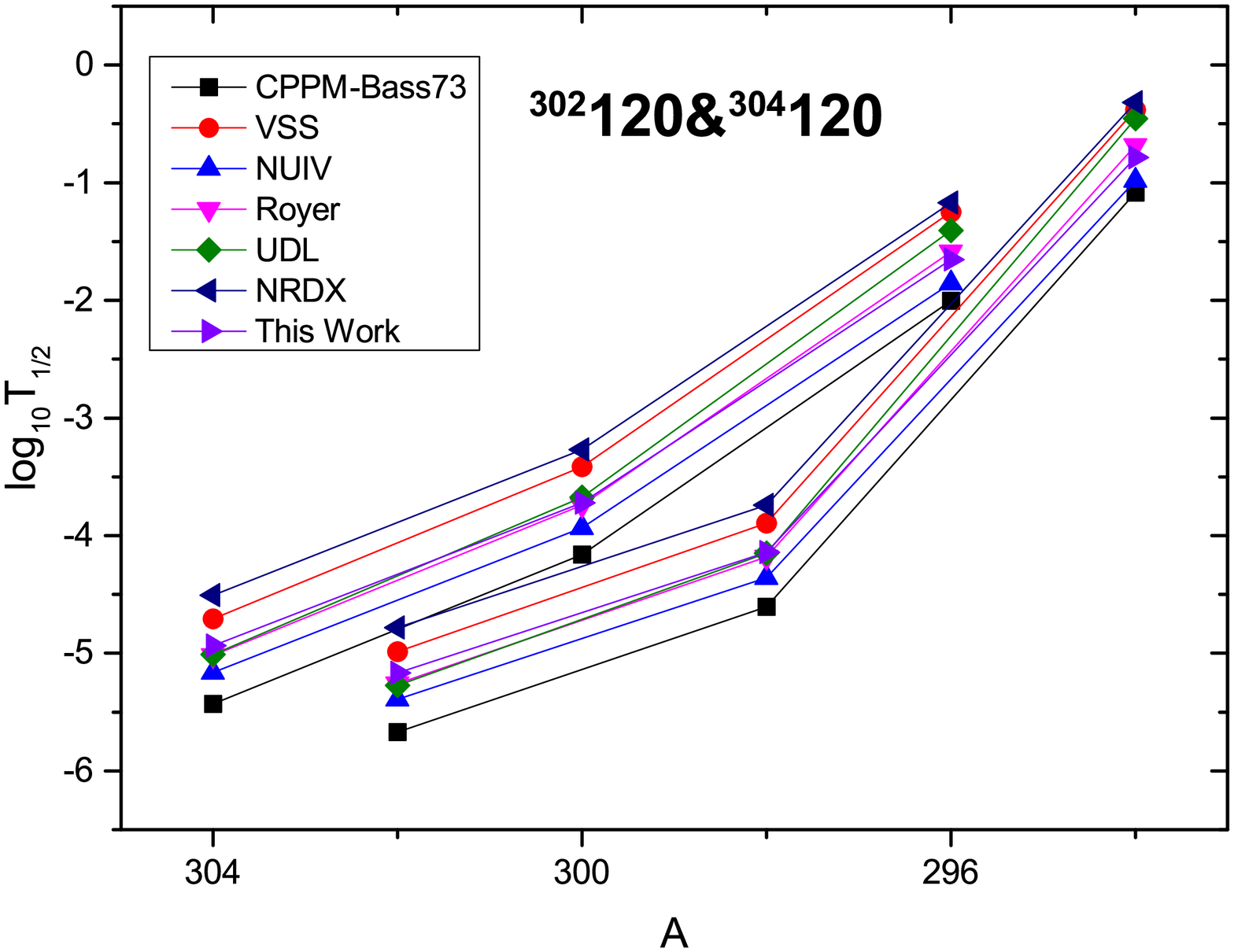}}
 \centerline{The case of $ ^{302}120$ and $ ^{304}120$}
\end{minipage}
\vfill
\begin{minipage}{0.48\linewidth}
 \centerline{\includegraphics[height=5.0cm,width=6.0cm]{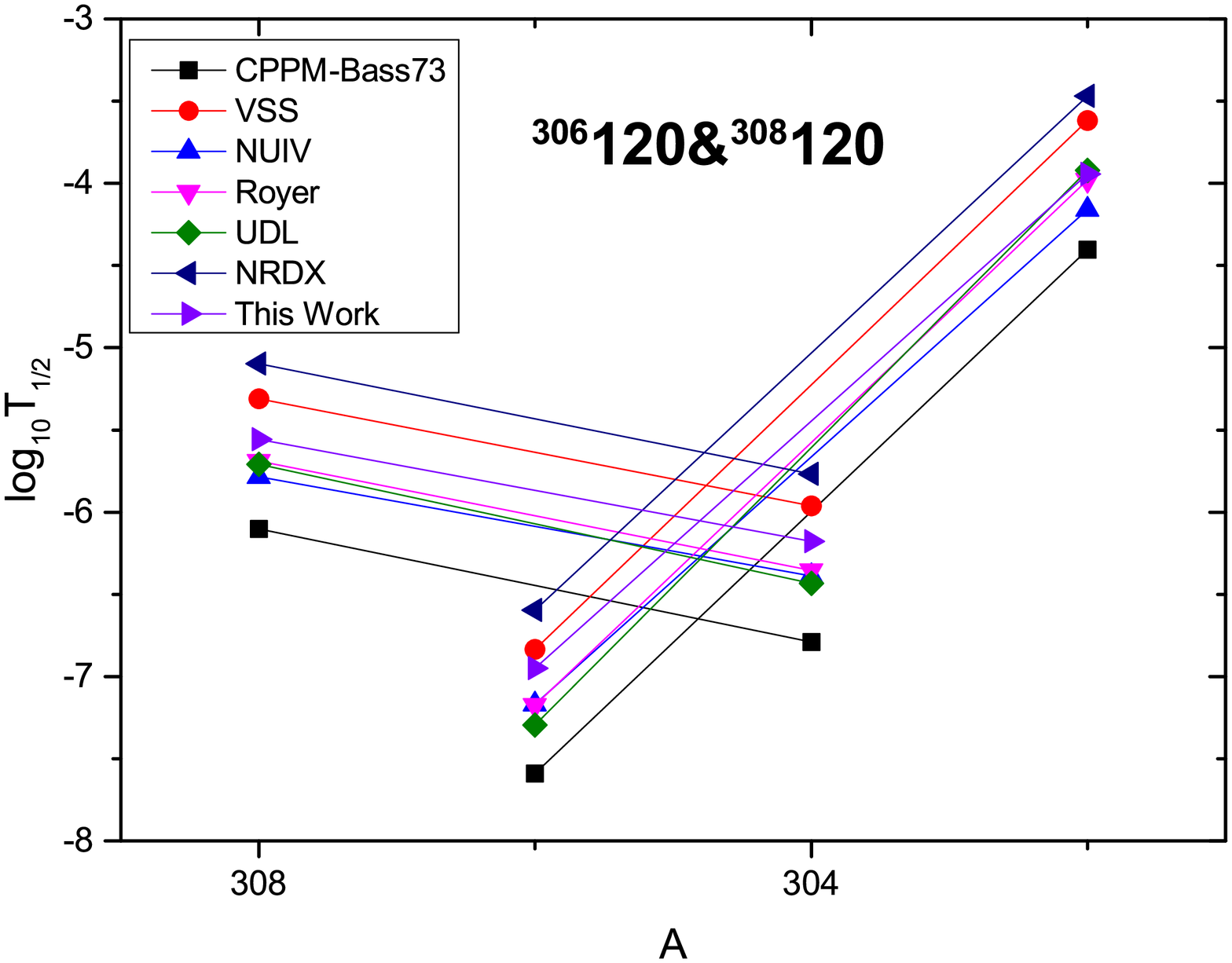}}
 \centerline{The case of $ ^{306}120$ and $ ^{308}120$}
\end{minipage}
\caption{The different cases of $\mathcal{\alpha}$ decay half-lives calculated by different theories. The abscissa A represents the mass of the nucleus, and the ordinate is the theoretical value of the $\mathcal{\alpha}$ decay half-life, the different color lines represent the calculations using different theoretical models.}
\label{fig1}
\end{figure}

For comparatively, we also systematically calculate the $\mathcal{\alpha}$ decay half-lives of even-even nuclei of proton numbers $Z=120$ and nuclei on their $\mathcal{\alpha}$ decay chain using Coulomb potential and Proximity potential model with proximity potential Bass73 formalism (CPPM-Bass73) \cite{BASS197445}, the Viola-Seaborg-Sobiczewski (VSS) empirical formula \cite{VIOLA1966741}, the Universal (UNIV) curve \cite{PhysRevC.83.014601, PhysRevC.85.034615}, Royer formula \cite{0954-3899-26-8-305}, the Universal decay law (UDL) \cite{PhysRevLett.103.072501, PhysRevC.80.044326} and the Ni-Ren-Dong-Xu (NRDX) empirical formula \cite{PhysRevC.78.044310}, respectively. The logarithmic forms of calculated $\mathcal{\alpha}$ decay half-lives are listed in Table \ref{table 7}. In this tables, the first two columns represent the parent nucleus of the $\mathcal{\alpha}$ decay and the $\mathcal{\alpha}$ decay energy, the next seven columns represent the theoretical $\mathcal{\alpha}$ decay half-lives calculated by CPPM-Bass73, VSS, UNIV, Royer, UDL, NRDX and our improved Gamow-like model denoted as ${\text{lg}{T^{\text{CPPM-Bass73}}_{1/2}}}$(s), ${\text{lg}{T^{\text{VSS}}_{1/2}}}$(s), ${\text{lg}{T^{\text{UNIV}}_{1/2}}}$(s), ${\text{lg}{T^{\text{Royer}}_{1/2}}}$(s), ${\text{lg}{T^{\text{UDL}}_{1/2}}}$(s), ${\text{lg}{T^{\text{NRDX}}_{1/2}}}$(s) and ${\text{lg}{T^{\text{This work}}_{1/2}}}$(s), respectively. The last column represents logarithmic form of the experimental $\mathcal{\alpha}$ decay half-lives taken from NUBASE2016 \cite{1674-1137-41-3-030001}. It can be seen from Table \ref{table 7} that for the $\mathcal{\alpha}$ decay of the same parent nuclear, the logarithmic form of theoretical $\mathcal{\alpha}$ decay half-life of all models are not much different, and the $\mathcal{\alpha}$ decay theoretical half-lives of the CPPM-Bass73 model is smaller than other models. For the parent nuclei with known experimental half-life, the maximum difference between the logarithmic forms of experimental half-life value and the logarithmic forms of theoretical half-life obtained from the model of this work is less than 0.65. To make a more intuitive comparison of these theoretical predictions, the theoretical half-life of $\mathcal{\alpha}$ decay calculated using this seven theoretical models are plotted in Fig. \ref{fig1}. In this figure, decay chains begin with an nucleus with a proton number $Z=120$, the nucleus at the end of each decay chain is spontaneous fission, and the decay of the remaining nucleus is $\mathcal{\alpha}$ decay. The X-axis represents the mass number in the corresponding $\mathcal{\alpha}$ decay chain, the Y-axis represents the logarithmic of the $\mathcal{\alpha}$ decay half-life.

\begin{table*}[!htbp]

\caption{Partial experimental data of $\mathcal{\alpha}$ decay and the predicted results as logarithmic forms of the theoretical values of $\mathcal{\alpha}$ decay half-live calculated by different methods of proton numbers $Z=120$ and nuclei on their $\mathcal{\alpha}$ decay chain.}
\label{table 7}
\renewcommand\arraystretch{1.6}
\setlength{\tabcolsep}{0.6mm}
\centering
\begin{footnotesize}
\centering
\begin{tabular}{cccccccccc}
Nucleus &$Q_{\alpha}$ (MeV)&${\text{lg}{T^{\text{CPPM-Bass73}}_{1/2}}}$(s)&${\text{lg}{T^{\text{VSS}}_{1/2}}}$(s)&${\text{lg}{T^{\text{UNIV}}_{1/2}}}$(s)&${\text{lg}{T^{\text{Royer}}_{1/2}}}$(s)&${\text{lg}{T^{\text{UDL}}_{1/2}}}$(s)&${\text{lg}{T^{\text{NRDX}}_{1/2}}}$(s)&${\text{lg}{T^{\text{This work}}_{1/2}}}$(s)&$\text{lg}{T^{\text{expt}}_{1/2}}$ (s) \\
 \hline
\noalign{\global\arrayrulewidth1pt}\noalign{\global\arrayrulewidth0.4pt} \multicolumn{10}{c}{\textbf{$^{296}120\to^{292}\text{Og}\to^{288}\text{Lv}\to^{284}\text{Fl}\to^{280}\text{Cn}\to^{276}\text{Ds}\to^{272}\text{Hs}\to^{268}\text{Sg}$}}\\
$^{	296	}$	120	&	13.187	&	-6.189	&	-5.613	&	-5.884	&	-5.774	&	-5.842	&	-5.398	&	-5.662	&	--	\\
$^{	292	}$	Og	&	12.015	&	-4.264	&	-3.662	&	-4.044	&	-3.842	&	-3.809	&	-3.516	&	-3.832	&	--	\\
$^{	288	}$	Lv	&	11.105	&	-2.698	&	-2.082	&	-2.527	&	-2.275	&	-2.165	&	-1.994	&	-2.326	&	--	\\
$^{	284	}$	Fl	&	10.666	&	-2.202	&	-1.568	&	-2.018	&	-1.767	&	-1.647	&	-1.515	&	-1.831	&	--	\\
$^{	280	}$	Cn	&	10.911	&	-3.471	&	-2.797	&	-3.183	&	-2.999	&	-2.975	&	-2.744	&	-3.006	&	--	\\
$^{	276	}$	Ds	&	10.976	&	-4.259	&	-3.555	&	-3.891	&	-3.76	&	-3.802	&	-3.508	&	-3.723	&	--	\\
$^{	272	}$	Hs	&	9.54	&	-1.077	&	-0.406	&	-0.823	&	-0.603	&	-0.477	&	-0.425	&	-0.678	&	--	\\
\noalign{\global\arrayrulewidth1pt}\noalign{\global\arrayrulewidth0.4pt} \multicolumn{9}{c}{\textbf{$^{298}120\to^{294}\text{Og}\to^{290}\text{Lv}\to^{286}\text{Fl}\to^{282}\text{Cn}\to^{278}\text{Ds}\to^{274}\text{Hs}$}}\\																					 
$^{	298	}$	120	&	12.9	&	-5.643	&	-5.032	&	-5.371	&	-5.231	&	-5.258	&	-4.826	&	 -5.148	&	--	\\
$^{	294	}$	Og	&	11.835	&	-3.889	&	-3.254	&	-3.688	&	-3.471	&	-3.408	&	-3.115	&	-3.474	&	-2.939	\\
$^{	290	}$	Lv	&	11.005	&	-2.482	&	-1.832	&	-2.319	&	-2.062	&	-1.932	&	-1.748	&	-2.12	&	-2.097	\\
$^{	286	}$	Fl	&	10.365	&	-1.431	&	-0.771	&	-1.28	&	-1.008	&	-0.833	&	-0.731	&	-1.097	&	-0.456	\\
$^{	282	}$	Cn	&	10.106	&	-1.375	&	-0.695	&	-1.186	&	-0.934	&	-0.776	&	-0.677	&	-1.014	&	--	\\
$^{	278	}$	Ds	&	10.31	&	-2.601	&	-1.882	&	-2.315	&	-2.122	&	-2.057	&	-1.862	&	-2.15	&	--	\\
\noalign{\global\arrayrulewidth1pt}\noalign{\global\arrayrulewidth0.4pt} \multicolumn{9}{c}{\textbf{$^{300}120\to^{296}\text{Og}\to^{292}\text{Lv}\to^{288}\text{Fl}\to^{284}\text{Cn}$}}\\																						 
$^{	300	}$	120	&	13.287	&	-6.461	&	-5.811	&	-6.13	&	-6.045	&	-6.116	&	-5.591	&	 -5.907&	--	\\
$^{	296	}$	Og	&	11.561	&	-3.279	&	-2.612	&	-3.109	&	-2.867	&	-2.759	&	-2.484	&	-2.895	&	--	\\
$^{	292	}$	Lv	&	10.775	&	-1.922	&	-1.243	&	-1.784	&	-1.51	&	-1.338	&	-1.168	&	-1.587	&	-1.602	\\
$^{	288	}$	Fl	&	10.065	&	-0.624	&	0.06	&	-0.506	&	-0.214	&	0.017	&	0.087	&	-0.325	&	-0.125	\\
\noalign{\global\arrayrulewidth1pt}\noalign{\global\arrayrulewidth0.4pt} \multicolumn{9}{c}{\textbf{$^{302}120\to^{298}{\text{Og}}\to^{294}\text{Lv}\to^{290}\text{Fl}$}}\\																						 
$^{	302	}$	120	&	12.878	&	-5.671	&	-4.986	&	-5.391	&	-5.259	&	-5.273	&	-4.781	& -5.166	&	--	\\
$^{	298	}$	Og	&	12.118	&	-4.607	&	-3.893	&	-4.358	&	-4.182	&	-4.148	&	-3.741	&	-4.144	&	--	\\
$^{	294	}$	Lv	&	10.451	&	-1.083	&	-0.379	&	-0.981	&	-0.683	&	-0.453	&	-0.319	&	-0.785	&	--	\\
\noalign{\global\arrayrulewidth1pt}\noalign{\global\arrayrulewidth0.4pt} \multicolumn{9}{c}{\textbf{$^{304}120\to^{300}\text{Og}\to^{296}\text{Lv}\to^{294}\text{Fl}$}}\\																	 
$^{	304	}$	120	&	12.745	&	-5.43	&	-4.71	&	-5.162	&	-5.019	&	-5.011	&	-4.509	&	 -4.937	&	--	\\
$^{	300	}$	Og	&	11.905	&	-4.162	&	-3.414	&	-3.935	&	-3.741	&	-3.672	&	-3.27	&	-3.719	&	--	\\
$^{	296	}$	Lv	&	10.777	&	-2.002	&	-1.248	&	-1.853	&	-1.588	&	-1.406	&	-1.172	&	-1.654	&	--	\\
\noalign{\global\arrayrulewidth1pt}\noalign{\global\arrayrulewidth0.4pt} \multicolumn{9}{c}{\textbf{$^{306}120\to^{302}\text{Og}\to^{298}\text{Lv}$}}\\			
$^{	306	}$	120	&	13.823	&	-7.59	&	-6.836	&	-7.169	&	-7.175	&	-7.296	&	-6.595	&	-6.949	&	--	\\
$^{	302	}$	Og	&	11.995	&	-4.404	&	-3.618	&	-4.16	&	-3.98	&	-3.92	&	-3.47	&	-3.944	&	--	\\
\noalign{\global\arrayrulewidth1pt}\noalign{\global\arrayrulewidth0.4pt} \multicolumn{9}{c}{\textbf{$^{308}120\to^{304}\text{Og}\to^{300}\text{Lv}$}}\\					
$^{	308	}$	120	&	13.036	&	-6.102	&	-5.309	&	-5.784	&	-5.689	&	-5.709	&	-5.096	&	-5.559	&	--	\\
$^{	304	}$	Og	&	13.104	&	-6.789	&	-5.96	&	-6.389	&	-6.354	&	-6.434	&	-5.769	&	-6.178	&	--	\\
 \hline
\end{tabular}
\end{footnotesize}

\end{table*}

From Fig. \ref{fig1}, we can clearly see that theoretical calculations of $\mathcal{\alpha}$ decay half-lives by different models of the same nucleus are different due to the model dependent. But all theoretically calculated $\mathcal{\alpha}$ decay half-life curves have the same trend. In order to intuitively compare different theories, we calculate the standard deviation $\Delta=\sqrt{\sum ({\text{lg}{T^{\text{expt}}_{1/2}}}-{\text{lg}{T^{\text{cal}}_{1/2}}})^2/n}$ between $\mathcal{\alpha}$ decay half-lives of calculations and experimental data of different theories in Table \ref{table 8}.
\begin{table}
\caption{$\Delta$ between $\mathcal{\alpha}$ decay half-lives of calculations and experimental data of different theories.}
\label{table 8}

\begin{tabular}{cccccccc}

\multicolumn{1}{c}{A of nucleus}&$\Delta_{CPPM-bass73}$&$\Delta_{VSS}$&$\Delta_{UNIV}$&$\Delta_{Royer}$&$\Delta_{UDL}$&$\Delta_{NRDX}$&$\Delta_{This -work}$\\
\hline
\\
298&0.817&	0.299&	0.656&	0.443&	0.360&	0.276&0.482
\\
300&0.419&	0.286&	0.299&	0.091&	0.212&	0.342&0.142
\\
\hline
\end{tabular}
\end{table}
We can clearly see that NRDX model reproduces experimental half lives well in superheavy region in case $^{298}120$,  Royer formula reproduces experimental half lives well in superheavy region in case $^{300}120$.  In particular, our calculations are sandwiched in other decay chains, and closed to the known experimental data for the $\mathcal{\alpha}$ half-lives, which shows that the model and calculated parameters of present work are believable.

\section{Summary}
In summary, we modify the Gamow-like model by considering the effects of screened electrostatic for Coulomb potential and the centrifugal potential and use this model systematically to study $\mathcal{\alpha}$ decay half-lives for $Z>51$ nuclei. In addition, we extend this model to the superheavy nuclei, and predict the half-lives of seven even-even nuclei with a proton number $Z=120$ and some un-synthesized nuclei on their $\mathcal{\alpha}$ decay chains. This work is useful for the future research of superheavy nuclei.
\label{section 4}

\section*{Acknowledgements}
This work is supported in part by the National Natural Science Foundation of China (Grant No. 11205083), the construct program of the key discipline in Hunan province, the Research Foundation of Education Bureau of Hunan Province, China (Grant No. 15A159 and 18A237), the Natural Science Foundation of Hunan Province, China (Grant No. 2015JJ3103 and No. 2015JJ2123), the Innovation Group of Nuclear and Particle Physics in USC, the double first class construct program of USC.

\section*{References}
%\bibliography{lixh-elsarticle-template}

\end{document}